\pdfoutput=1

\documentclass[prl,twocolumn,showpacs,preprintnumbers,amsmath,amssymb,
superscriptaddress,nofootinbib]{revtex4-1}
\usepackage{epsfig}
\usepackage{graphicx}
\usepackage{color}

\newcommand{\smallminus}{{\rm\rule[2.4pt]{6pt}{0.65pt}}}
\newcommand{\smallplus}{\hspace{0.5pt}\text{{\small+}}\hspace{-0.5pt}}
\newcommand{\mi}{\smallminus}
\newcommand{\psl}{\smallplus}

\begin{document}

\preprint{}

\title{Cutting Deep Into The Amplituhedron}

\author{Nima Arkani-Hamed}
\affiliation{School of Natural Sciences, Institute for Advanced Study, Princeton, NJ, USA}

\author{Cameron Langer}
\affiliation{Center for Quantum Mathematics and Physics (QMAP), University of California, Davis, CA, USA}

\author{Akshay Yelleshpur Srikant}
\affiliation{Department of Physics, Princeton University, NJ, USA}

\author{Jaroslav Trnka}
\affiliation{Center for Quantum Mathematics and Physics (QMAP), University of California, Davis, CA, USA}

%\date{\today}

\begin{abstract}
In this letter we compute a canonical set of cuts of the integrand for MHV amplitudes in planar ${\cal N}=4$ SYM, where all internal propagators are put on-shell. These ``deepest cuts" probe the most complicated Feynman diagrams and on-shell processes that can possibly contribute to the amplitude, but are also naturally associated with remarkably simple geometric facets of the amplituhedron. The recent reformulation of the amplituhedron in terms of combinatorial geometry directly in the kinematic (momentum-twistor) space plays a crucial role in understanding this geometry and determining the cut. This provides us with the first non-trivial results on scattering amplitudes in the theory valid for arbitrarily many loops and external particle multiplicities. \end{abstract}

\maketitle

%%%%%%%%%%%%%%%%%
\section{Introduction}
%%%%%%%%%%%%%%%%%

The past decade has revealed a variety of surprising mathematical and physical structures underlying particle scattering amplitudes, providing, with various degrees of completeness, reformulations of this physics where the normally foundational principles of locality and unitarity are derivative from ultimately combinatoric-geometric origins. An example is the amplituhedron \cite{Arkani-Hamed:2013jha}, a geometric picture for scattering amplitudes in planar ${\cal N}=4$ SYM theory. All tree-level amplitudes and loop integrands in this theory correspond to the differential forms with logarithmic singularities on  boundaries of the amplituhedron geometry. The original definition was based on a generalization of the positive Grassmannian, centrally connected to on-shell diagrams \cite{ArkaniHamed:2012nw} and loop level recursion relations \cite{ArkaniHamed:2010kv}. More recently, a more intrinsic definition of the amplituhedron was found \cite{Arkani-Hamed:2017vfh}, directly in momentum-twistor space, using certain topological notions--winding numbers and sign flip patterns--associated with projections of the momentum-twistor data. The amplituhedron has been extensively studied from many angles including mathematical aspects \cite{Lam:2014jda,Karp:2016uax,Galashin:2018fri}, positive geometry and volume interpretation \cite{Arkani-Hamed:2014dca,Arkani-Hamed:2017tmz}, triangulations \cite{Arkani-Hamed:2013kca,Franco:2014csa,Galloni:2016iuj,Rao:2018uta}, connections to on-shell diagrams  \cite{Bai:2014cna,Bai:2015qoa} and geometric structures in the final amplitudes \cite{Dixon:2016apl,Dennen:2016mdk}. Some early steps in extending this circle of ideas well beyond the planar ${\cal N}=4$ SYM have been taken in \cite{Bern:2015ple,Arkani-Hamed:2017mur,Salvatori:2018aha}.

There has also been an ongoing effort to use the amplituhedron picture to make all-loop order predictions for loop integrands. This effort was initiated in \cite{Arkani-Hamed:2013kca}, which calculated certain all-loop order cuts for four point amplitudes which were impossible to obtain using any other methods. In this letter, we go much further and use the new topological definition of the amplituhedron \cite{Arkani-Hamed:2017vfh} to calculate a particular cut of $n$-point MHV amplitudes to all loops. This cut places on-shell internal propagators which are arbitrarily deep in the interior of contributing Feynman diagrams. Thus, we aptly refer to this as the ``deepest cut." It appears hopelessly difficult  to calculate this cut using standard unitarity-based methods, as almost all diagrams contribute as we increase the loop order. However, we will show that the new topological formulation of the amplituhedron allows us to easily understand the geometry of the facet associated with this cut and leads to a strikingly simple, one-line expression for the cut valid  to all loops and all multiplicities.  This is the first calculation giving us non-trivial access to the regime of arbitrarily large loop order and particle multiplicities in the theory. 

%%%%%%%%%%%%%%%%%%%
\section{Amplitudes from sign flips}
%%%%%%%%%%%%%%%%%%%

The original definition of the amplituhedron refers to the auxiliary space of extended kinematical variables constrained by positivity conditions. Recently, an equivalent definition was provided directly in the momentum twistor space using the conditions on {\it sign flips} \cite{Arkani-Hamed:2017vfh}. 

For the $n$-point amplitude, the kinematics is given by $n$ momentum twistors $Z_a^I$, $a=1,2,\dots,n$, $I=1,\dots,4$. The $SL(4)$ dual conformal symmetry acts on the $I$ index.  We define the $SL(4)$ invariants $\langle abcd\rangle \equiv \epsilon_{IJKL}Z_a^IZ_b^JZ_c^KZ_d^L$. The space for the N$^{k}$MHV amplituhedron is described by the set of $Z_a$ for which all $\langle i\,i\psl1\,j\,j\psl1\rangle \geq 0$ (with twisted positivity for $(-1)^{k+1}\langle n\,1\,j\,j\psl1\rangle>0$), and the following sequence
\begin{equation}
\{\langle 123\,i\rangle\}\,\,\,\mbox{for}\,\,i=4,\dots,n\,\,\,\,\mbox{has $k$ sign flips}.
\label{Flip1}
\end{equation}
For example, for $n=6$ the sequence has three terms,
\begin{equation*}
\{\langle 1234\rangle, \langle1235\rangle, \langle1236\rangle\},
\label{Flip2}
\end{equation*}
and possible sign sequences $\{+++\}$, $\{++-\}$, $\{+--\}$ and $\{+-+\}$. The first sequence corresponds to $k=0$, the next two to $k=1$ and the last to $k=2$ kinematics. In general the $k=0$ MHV amplitude has the sign pattern $\{++\dots+\}$ and zero sign flips, while for higher $k$ we have various sign patterns which have $k$ sign flips in total.

At loop level, in addition to the external momentum twistors $Z_a$, we also have lines $(AB)_\alpha$ corresponding to loop momenta. For each line we can write $(AB)_\alpha=A_\alpha B_\alpha$ where $A_\alpha$, $B_\alpha$ are two points on that line. The line $(AB)_\alpha$ is in the one-loop amplituhedron if all $\langle (AB)_\alpha\,i\,i+1\rangle>0$ (again with twisted positivity $(-1)^{k+1}\langle (AB)_\alpha\, n1\rangle>0$) and if the sequence 
\begin{equation}
\{\langle (AB)_\alpha \,1\,i\rangle\}\,\,\,\mbox{for}\,\,i=2,\dots,n\,\,\,\mbox{has $(k+2)$ sign flips}.
\label{ABpos}
\end{equation}
The collection of $\ell$ lines, $\{(AB)_\alpha\}$ is in the $\ell$-loop amplituhedron if each line satisfies (\ref{ABpos}) and in addition the mutual positivity conditions (independent of $n$, $k$)
\begin{equation}
\langle (AB)_\alpha\,(AB)_\beta \rangle>0 \quad \mbox{for all $\alpha,\beta=1,\dots,\ell$}.
\end{equation}
The $n$-point N$^k$MHV $\ell$-loop integrand is given by a degree $4(k + \ell)$ differential form on $\left\{Z_a,(AB)_\alpha \right\}$ space, determined by the property of having logarithmic singularities on boundaries of the intersection of this space with a canonical $(4 \times k)$-dimensional affine subspace in the configuration space of momentum twistors $\left\{Z_a\right\}$.  

The amplituhedron geometry for the MHV case when $k=0$ is especially simple. The space is simply that of $\ell$ lines 
$(AB)_{\alpha}$ in momentum twistor space. The sign-flip conditions  (\ref{ABpos}) can be rewritten in an apparently different but equivalent form, in terms of inequalities on each $(AB)_{\alpha}$:
\begin{equation}
\langle (AB)_\alpha \, (i\mi1\,i\,i\psl1)\cap(j\mi1\,j\,j\psl1)\rangle>0\quad\mbox{for all $i,j$}.
\label{ABpos2}
\end{equation}
This condition is the same for every loop--we can say that this just demands that each $(AB)_\alpha$ lives in the one-loop amplituhedron. The interaction between different loops is then captured by the mutual positivity properties $\langle (AB)_\alpha (AB)_\beta\rangle >0$.

%
%where the intersection of two planes is a line, $(abc)\cap(def) = Z_aZ_b \langle cdef\rangle + Z_bZ_c\langle adef\rangle + Z_cZ_a\langle bdef\rangle$. 

%%%%%%%%%%%%%%%%%%%
\section{Definition of the deepest cut}
%%%%%%%%%%%%%%%%%%%

In \cite{Arkani-Hamed:2013kca} we focused on cuts where ${\cal L}_\alpha$ passed through $Z_i$ or cut lines $Z_iZ_{i{+}1}$. Here we consider an opposite case where none of the external propagators $\langle (AB)_\alpha  i\,i{+}1\rangle$ are cut but all internal propagators are on-shell
\begin{equation}
\langle (AB)_\alpha \,(AB)_\beta \rangle=0\qquad \mbox{for all $\alpha,\beta=1,\dots,\ell$}.
\label{Cut}
\end{equation}

Prior to any detailed investigation, the geometry of the amplituhedron makes an amazing prediction for the structure of this cut. Owing to the trivialization of the mutual positivity by setting all the $\langle (AB)_{\alpha} (AB)_{\beta} \rangle\to 0$, the only remaining constraint of the geometry is that all the lines $(AB)_\alpha$ live in the one-loop amplituhedron! This leads us to expect that the all-loop geometry should be expressible as $\ell$ independent copies of the one-loop geometry, associated with a formula for the cut with the structure of a product over independent pieces determined by this one-loop problem,  We will see this expectation borne out perfectly in our analysis.

It is easy to show that we have to impose $2\ell-3$ conditions in order to satisfy (\ref{Cut}). There are two solutions to this problem, each with a different geometrical meaning: in the first solution, {\it all-in-point}, all lines intersect in a common point $A$ while in the second solution, {\it all-in-plane}, all lines lie on the same plane $P$.
$$
\includegraphics[scale=.48]{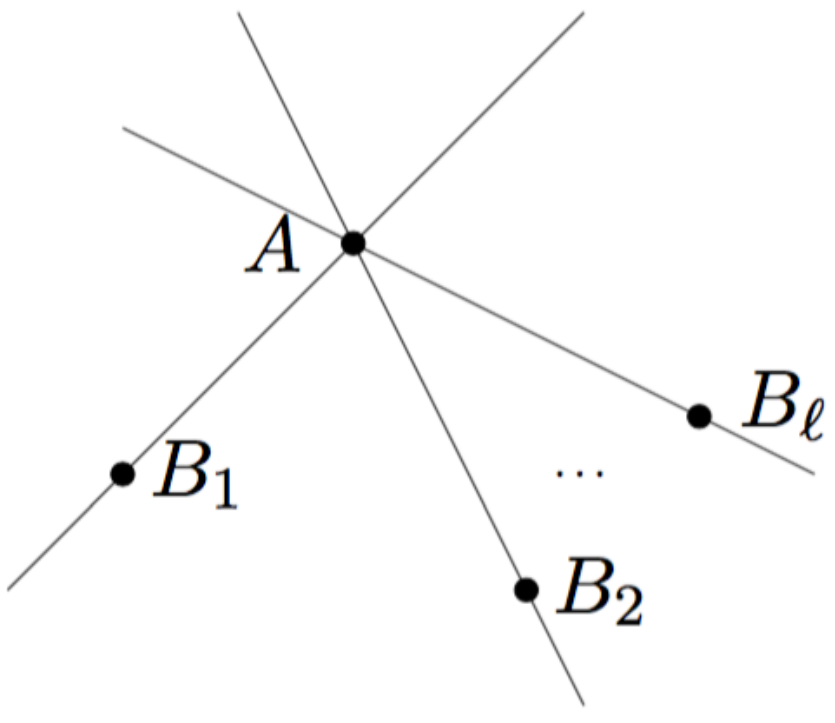}\,\,\,
\includegraphics[scale=.48]{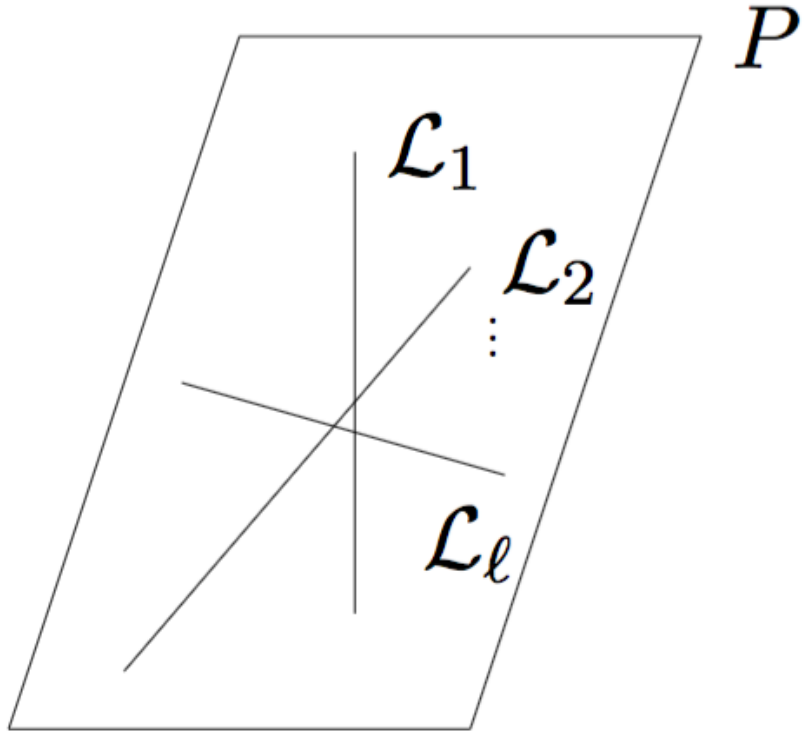}
$$
These configurations are mapped into each other by the usual projective duality interchanging points and planes, which also reflects parity; had we been discussing $\overline{\rm MHV}$ amplitudes the cuts associated with the pictures would be reversed. In the first solution the lines $(AB)_\alpha$ can be parametrized 
\begin{equation}
(AB)_\alpha = AB_\alpha,
\end{equation}
where $B_\alpha$ and $A$ are arbitrary. The common intersection point $A$ has three degrees of freedom while each $B_\alpha$ has two degrees of freedom as the geometry only depends on the lines $AB_\alpha$. As a result, the configuration is $(2\ell+3)$- dimensional. We now have a simple geometry problem. Given a point $A$, we want to identify all the lines $(AB)$ passing through $A$, which lie in the one-loop amplituhedron. This carves out some subset in the two-dimensional space of possible points $B$. The shape of this region can change as we move around in $A$ space. Thus we are led to find a joint ``triangulation" in $A,B$ space, specified by breaking up the three-dimensional $A$ space into regions for which the corresponding two-dimensional geometry in $B$ has a uniform shape. The regions in $B_\alpha$ space are the same for all $\alpha$. Thus for each such piece we take the product of the $A$-form and the $B_\alpha$ forms, and then sum over all terms in the triangulation. Because there are no mutual positivity conditions left for the given point $A$, the form in $B_\alpha$ is given by a simple product. The final result can then be written as
\begin{equation}
\Omega_n^{(\ell)} = \sum_{j} \omega_j^{(n)}(A) \wedge \prod_{\alpha=1}^\ell \kappa_j^{(n)}(B_\alpha).
\label{form1}
\end{equation}
In the second solution all lines $(AB)_\alpha$ lie in the same plane $P$ and we denote them 
\begin{equation}
(AB)_\alpha={\cal L}_\alpha.\
\end{equation}
The plane $P$ has three degrees of freedom while each line ${\cal L}_\alpha$ has two degrees of freedom, which is $2\ell+3$ in total.  Again, we imagine fixing the plane $P$ and looking for all the lines ${\cal L}$ in $P$ that lie in the one-loop amplituhedron. This breaks up $P, {\cal L}$ space into pieces where, given $P$ in a certain region,  the corresponding geometry in ${\cal L}$-space has a uniform shape. 
The differential form for this space can be written as
\begin{equation}
\widetilde{\Omega}_n^{(\ell)} = \sum_j \widetilde{\omega}_j^{(n)}(P) \wedge \prod_{\alpha=1}^\ell \widetilde{\kappa}_j^{(n)}(
{\cal L}_\alpha).
\end{equation}
The various differential forms $\omega(A), \kappa(B_\alpha), \omega(P)$ and $\kappa({\cal L})$ are proportional to universal measure factors $\mathrm{d} \mu_A, \mathrm{d} \mu_B, \mathrm{d} \mu_P, \mathrm{d} \mu_{{\cal L}}$ associated with the free points, planes and lines characterizing the geometries. 
The measures for the point $A$ and points $B_\alpha$ are
\begin{align}
\mathrm{d}\mu_A &= \langle A\,\mathrm{d}^3A\rangle \equiv \epsilon^{IJKL} A^I\,\mathrm{d}A^J \wedge \mathrm{d}A^K \wedge \mathrm{d}A^L, \nonumber\\
\mathrm{d}\mu_B &= \langle AB\,\mathrm{d}^2B\rangle \equiv \epsilon^{IJKL} A^I B^J\,\mathrm{d}B^K \wedge \mathrm{d}B^L.
\end{align}
For the plane $P\equiv P^{IJK}$, we can define $P^{IJK} \equiv \epsilon^{IJKL} p_L$ and then the measure for $P$ is 
\begin{equation}
\mathrm{d}\mu_P = \langle p\,\mathrm{d}^3p\rangle \equiv \epsilon^{IJKL} p_I\,\mathrm{d}p_J\wedge \mathrm{d}p_K \wedge \mathrm{d}p_L. 
\end{equation}
The plane $P$ can be parametrized using three points $p_i^I$, $i=1,2,3$ up to a $GL(3)$ transformation on the $i$ index. Then the line $(AB)^{IJ}$ is related to the line ${\cal L}_k$ on $P$ as
\begin{equation}
(AB)^{IJ} = \epsilon^{ijk} (p_ip_j)^{IJ} {\cal L}_k\label{ABL},
\end{equation}
where the $\epsilon^{ijk}$ acts on the labels of points $p_i$ on the plane $P$. Finally, the measure of the line ${\cal L}$ is
\begin{equation}
\mathrm{d}\mu_{{\cal L}} = \langle {\cal L} {\cal L}\,\mathrm{d}^2{\cal L}\rangle \equiv \epsilon^{ijk} {\cal L}_i\,\mathrm{d}{\cal L}_j\wedge \mathrm{d}{\cal L}_k
\end{equation}

The richness of the deepest cut is revealed when compared to the amplitude written as a sum of Feynman integrals. Contributing diagrams must have at least $2\ell-3$ internal propagators. Ladder diagrams with only $\ell-1$ propagators are irrelevant while other diagrams with more internal propagators contribute such as 
$$
\includegraphics[scale=.41]{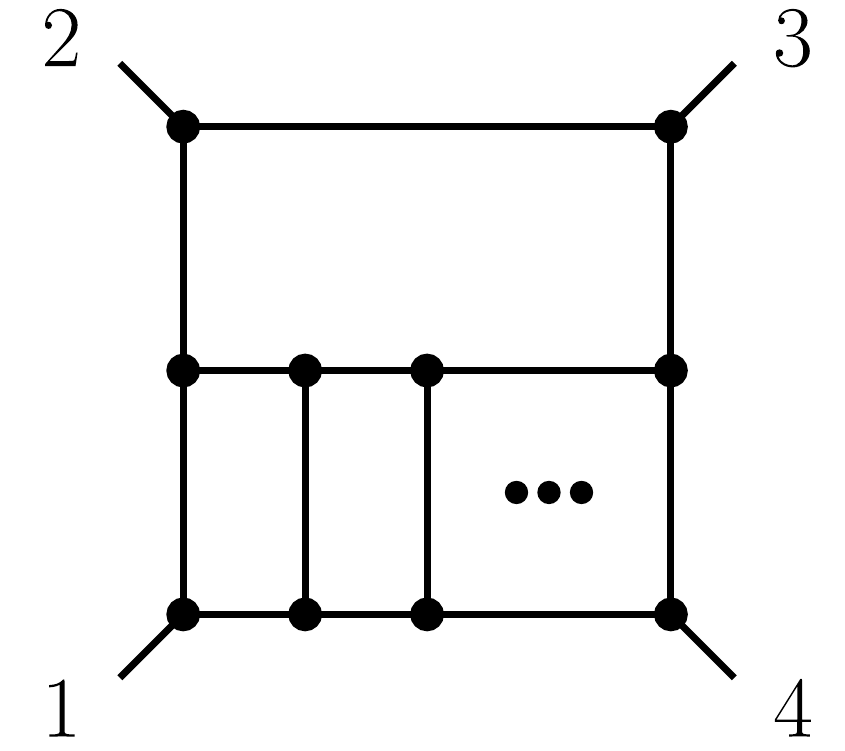}\,\,\,\,
\includegraphics[scale=.41]{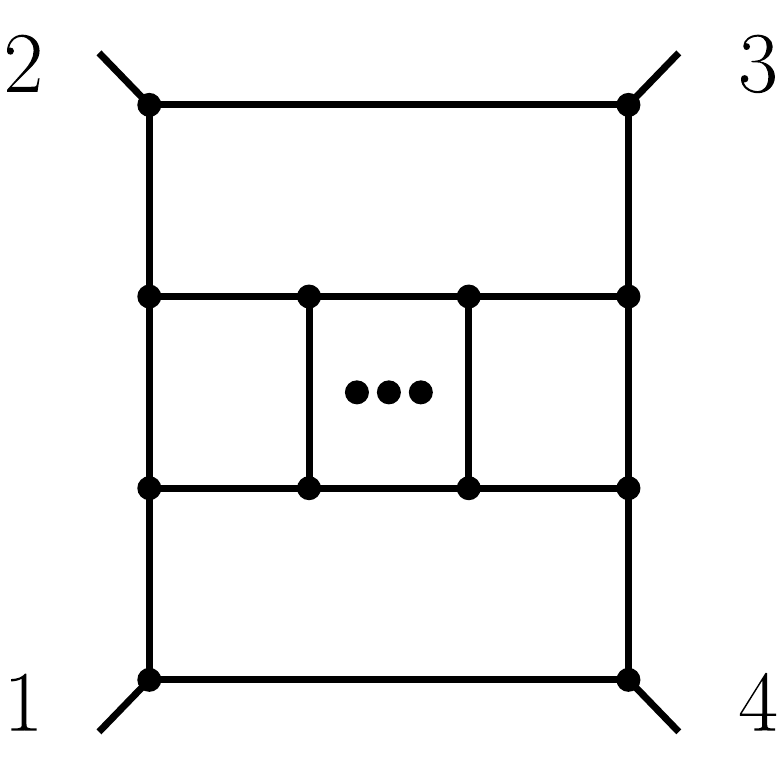}
$$
The first diagram has $2\ell{-}3$ propagators and is exactly borderline while the second more complicated diagram has $3\ell{-}7$ propagators. More internal propagators means more complicated cut structure and singularities which results in more complicated analytic structure and branch cuts after performing the loop integration. It is exactly these most complicated diagrams which contribute on the deepest cut.

%%%%%%%%%%%%%%%%%%%
\section{Four point case}
%%%%%%%%%%%%%%%%%%%

We now proceed to solving the geometry problem for the four point case where all lines pass through the same point $A$. The positivity of external data at four points is a single condition $\langle 1234\rangle>0$ while the sign flip conditions for lines ${\cal L}_\alpha = AB_\alpha$ turn into the set of inequalities
\begin{align}
&\langle AB_\alpha 12\rangle>0,\,\langle AB_\alpha 23\rangle>0,\,\langle AB_\alpha 34\rangle>0,\,\langle AB_\alpha 14\rangle>0\nonumber\\
&\hspace{2cm} \langle AB_\alpha 13\rangle<0,\,\langle AB_\alpha 24\rangle<0.
\label{AB4}
\end{align}
As there are no mutual inequalities between different lines $AB_\alpha$ in this problem once we solve the inequalities for one $B$ it is automatically solved for all of them, and the differential form is given by the simple product (\ref{form1}). As all inequalities involve the point $A$ we can project through that point to a two-dimensional plane which contains projected points $Z_1'$, $Z_2'$, $Z_3'$, $Z_4'$ and $B_k'$
\vspace{-0.3cm}
$$
\includegraphics[scale=.4]{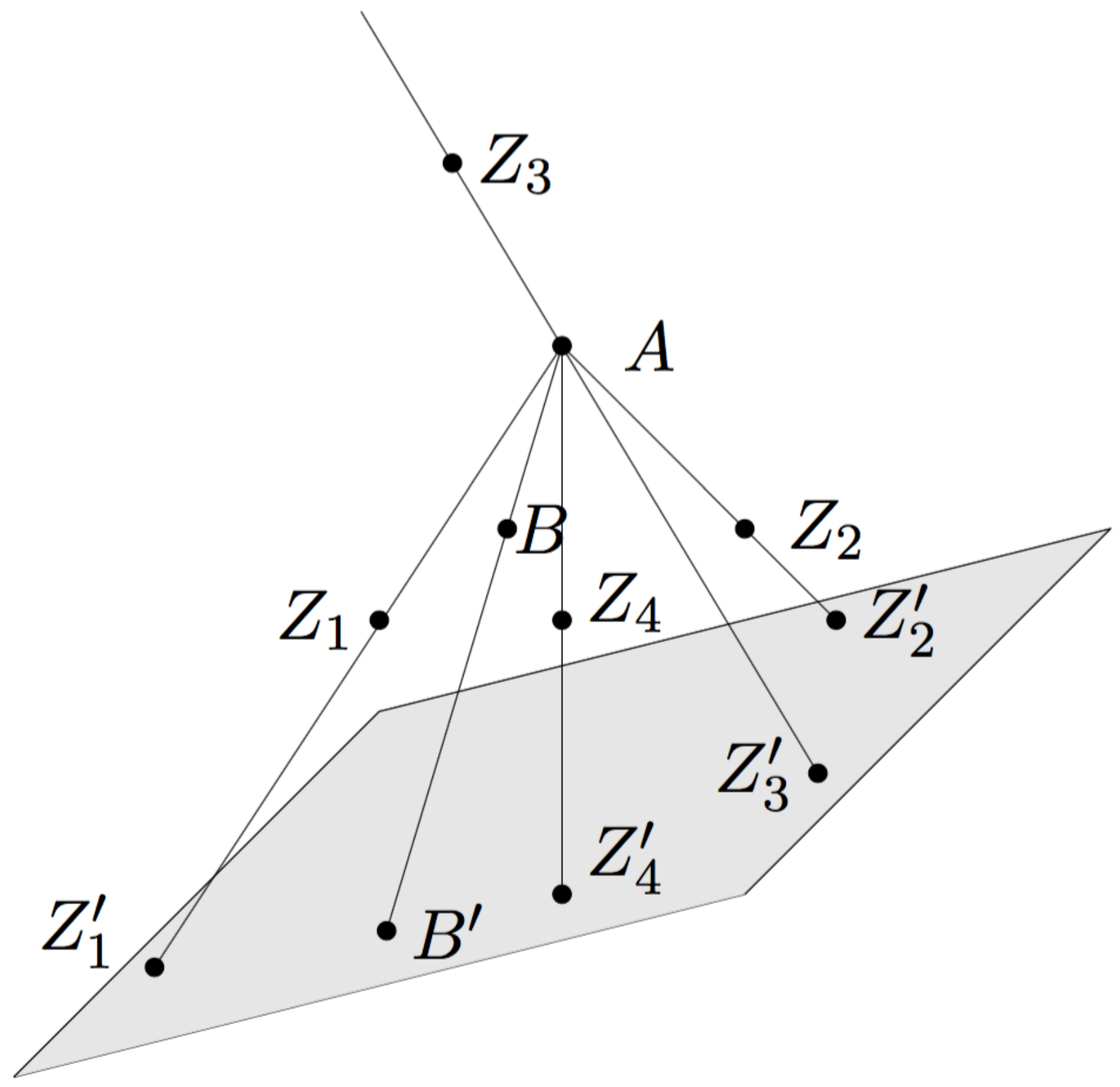}
$$
\vspace{-0.7cm}

\noindent We omit the primes in the following. These inequalities cut out an allowed region for $A$. The task is to triangulate this region such that for each term in the triangulation, there is a corresponding allowed region for $B$ whose shape does not change within this $A$ region. For the four point case the situation is quite simple: $A$ can be written as linear combination $A=c_1Z_1+c_2Z_2+c_3Z_3+c_4Z_4$ and the choice of signs of $c_j$ are equivalent to the choice of signs of the set,
\begin{equation}
\{\langle A123\rangle, \langle A124\rangle, \langle A134\rangle, \langle A234\rangle\}.
\label{Areg}
\end{equation}
There are $2^4=16$ possible choices corresponding to 16 $A$-regions for which we have to find the $B$-geometry. The two-dimensional configuration of the projected $Z_i$ points must respect given $\langle Aijk\rangle$ inequalities. In particular, $\langle Aijk\rangle>0$ if the triangle $(ijk)$ made of points $Z_i$, $Z_j$, $Z_k$ is oriented clockwise. The other sign $\langle Aijk\rangle<0$ corresponds to counterclockwise orientation. As a result, we get two types of configurations,
\vspace{-0.1cm}
$$
\includegraphics[scale=.32]{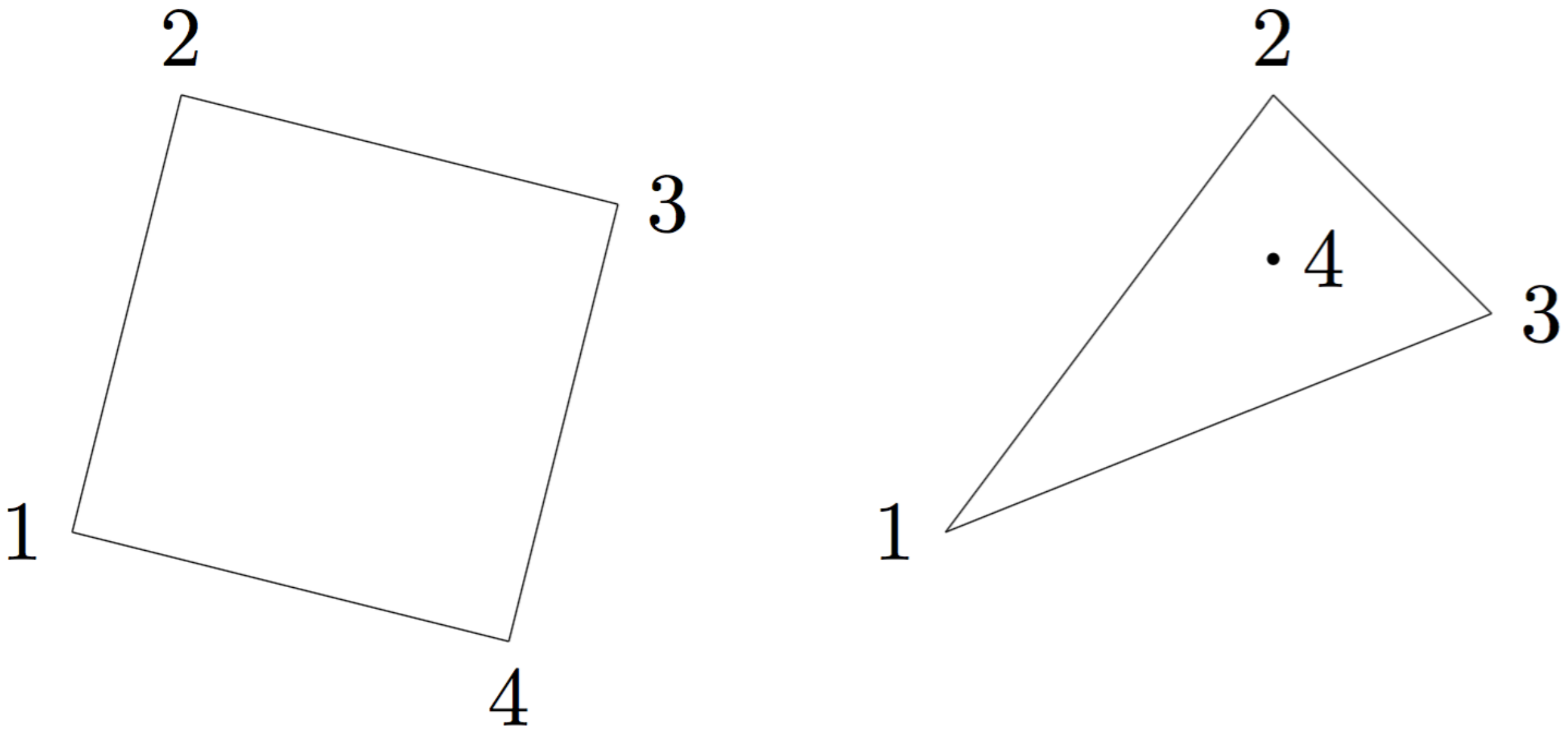}
$$
\vspace{-0.65cm}

where the first corresponds to $\{+,+,+,+\}$ and the second to $\{+,+,-,+\}$. Adding all possible permutations of $Z_i$ gives us eight quadrilateral configurations and eight triangle configurations with one point inside. In the second step we have to find the space of all $B$ points which satisfy inequalities (\ref{AB4}). Geometrically, the inequality $\langle ABij\rangle>0$ means that $B$ must be on the right side of the line $Z_iZ_j$ or alternatively the triangle $(Bij)$ must be oriented clockwise, for $\langle ABij\rangle<0$ counterclockwise. The first configuration has no allowed $B$-region while for the second we get
\vspace{-0.3cm}
$$
\includegraphics[scale=.48]{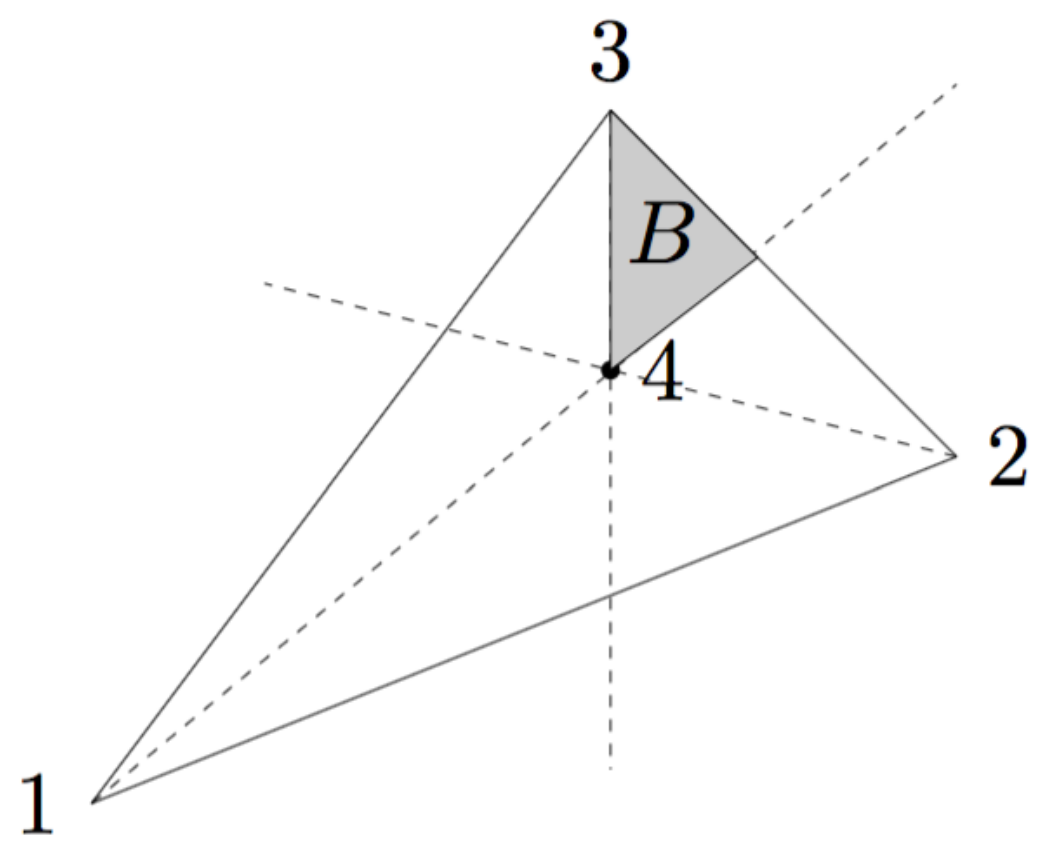}
$$
\vspace{-0.6cm}

which is a triangle given by the lines $(23)(34)(14)$. There are three more $A$-configurations which give non-empty $B$-regions. All these regions are triangles bounded by lines $(12)(23)(34)$, $(12)(34)(14)$ and $(12)(23)(14)$.

Since the $A$ and $B$ geometry are factorized, so are the corresponding logarithmic volume forms. For the $A$-part for all possible signs in (\ref{Areg}) the boundaries are the same, therefore the form is given for all possible regions by $(-1)^K\omega(A)$ where
\begin{equation}
\omega(A) = \frac{\mathrm{d}\mu_A\, \langle 1234\rangle^3}{\langle A123\rangle\langle A124\rangle\langle A134\rangle\langle A234\rangle},
\label{Aform}
\end{equation}
and $K$ is the number of minus signs in (\ref{Areg}). The $B$-forms depend on the shape of the allowed region in the two-dimensional plane. The triangle in the picture above is bounded by the lines $(23)$,$(34)$,$(14)$ and the form is
\begin{equation}
\omega(B) = \frac{\mathrm{d}\mu_B \langle A134\rangle \langle A234\rangle}{\langle AB23\rangle\langle AB34\rangle\langle AB14\rangle}.
\end{equation}
If the configuration was more complicated e.g., a pentagon, we would have to decompose it into triangles but that does not happen in the four point case. Putting all pieces together we can write the result as 
\begin{equation}
\Omega_4^{(\ell)} =\omega(A)\wedge\kappa(B),
\label{Omeg1}
\end{equation}
where $\omega(A)$ is given by (\ref{Aform}) and $\kappa(B)$ is a sum of four triangles with signs coming from different $A$-regions,
\begin{equation}
\hspace{-0.2cm} \kappa(B) = \sum_{j=1}^4 (-1)^j \prod_\alpha \frac{\mathrm{d}\mu_B\, \langle A\,j\mi1\,j\,j\psl1\rangle\langle Aj\,j\psl1\,j\psl2\rangle}{\langle AB_\alpha j\mi1\,j\rangle\langle AB_\alpha \,j\,j\psl1\rangle\langle AB_\alpha j\psl1\,j\psl2\rangle}
\end{equation}
where the sum over $j$ includes the cyclic twist: $n+k\rightarrow k$. 

In the four point case the all-in-plane solution can be extracted from the all-in-point solution. Instead of the common point $A$ we have a common plane $P=(P_1P_2P_3)$ on which all ${\cal L}_\alpha$ live. The space of $P$ planes is bounded by points $Z_1$, $Z_2$, $Z_3$, $Z_4$ and the form in $P$ up to a sign is given by 
\begin{equation}
\widetilde{\omega}(P) = \frac{\mathrm{d}\mu_P\,\langle 1234\rangle}{\langle P1\rangle\langle P2\rangle\langle P3\rangle\langle P4\rangle}
\end{equation}
On each plane $P$ we have two-forms on the space of lines ${\cal L}_\alpha$. The triangulation is analogous to the all-in-line case and the result can be written as $\widetilde{\Omega}_4^{(\ell)} = \widetilde{\omega}(P) \wedge \widetilde{\kappa}(AB)$ where
\begin{equation}
\hspace{-0.3cm}\widetilde{\kappa} = \sum_{j=1}^4 (-1)^j \prod_{\alpha=1}^\ell \frac{\mathrm{d}\mu_{{\cal L}_\alpha}\,\langle j{-}1jj{+}1j{+}2\rangle\langle Pj\rangle\langle Pj\psl1\rangle}{\langle (AB)_\alpha j\mi1\,j\rangle\langle (AB)_\alpha\,j\,j\psl1\rangle\langle (AB)_\alpha j\psl1\,j\psl2\rangle},
\end{equation}
where $(AB)_\alpha$ are related to the lines ${\cal L}_\alpha$ restricted to $P$ via (\ref{ABL}). Note that $\langle j{-}1jj{+}1j{+}2\rangle = (-1)^j \langle 1234\rangle$.

%%%%%%%%%%%%%%%%%%%
\section{Higher point formulas} 
%%%%%%%%%%%%%%%%%%%

At four points the original amplituhedron picture is identical to the new sign flip definition. However, for higher point MHV amplitudes while still equivalent the sign flip picture is much more suitable for actually solving the geometry. Here we provide the final $n$-point expressions for the residues on the deepest cut for both solutions; detailed derivations and a number of results for further non-trivial cuts will be provided in \cite{prep}. The basic strategy is the same as in the four point case: for the all-in-point solution, triangulate the $A$-space and find the corresponding $B_\alpha$ geometry. The main difference is that the boundaries of the different pieces in the triangulation of the $A$-space are now different.  The result can be schematically drawn as the tetrahedron $\times$ polygon,
$$
\includegraphics[scale=0.44]{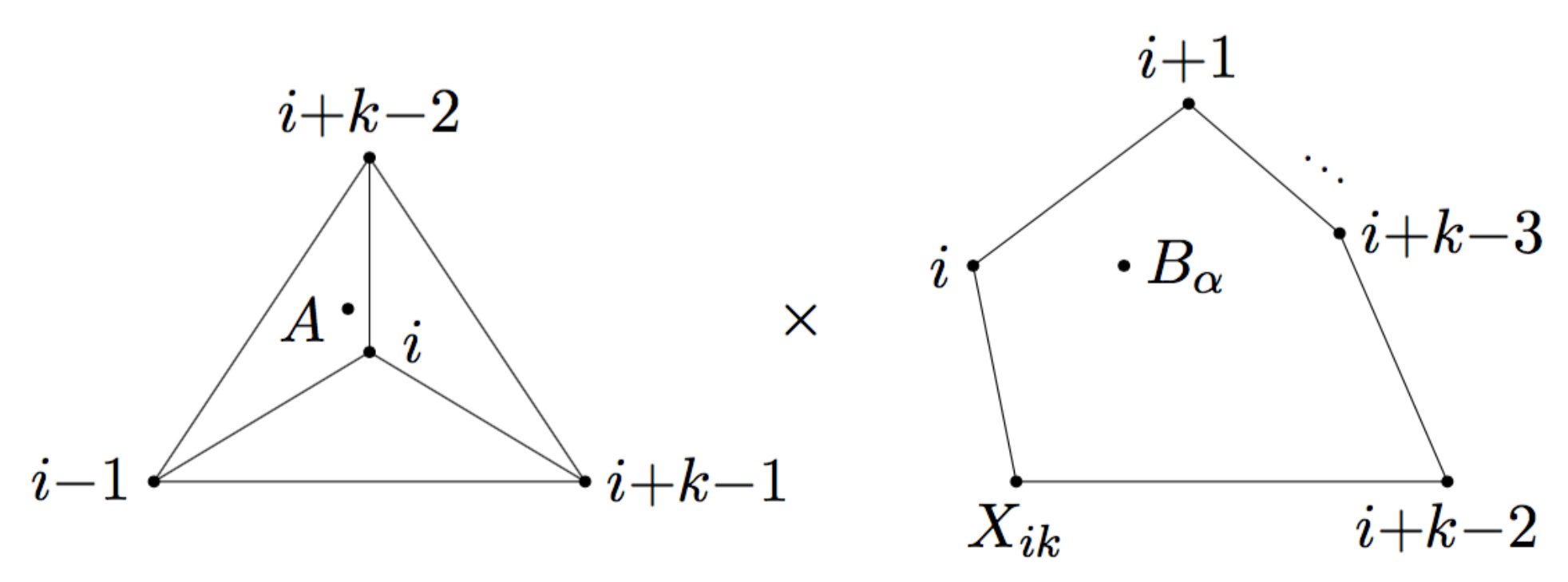}
$$
\vspace{-0.6cm}

\noindent where the point $X_{ik}\equiv (i{-}1i){\cap}(A,i{+}k{-}2,i{+}k{-}1)$. The expression for the integrand takes the product form 
\begin{widetext}
\vspace{-0.15cm}
\begin{equation}
\Omega_n^{(\ell)}=\sum_{k=1}^{n-1}\sum_{i=1}^n\,[i{-}1,i,i{+}k{-}2,i{+}k{-}1]\wedge\prod_{\alpha=1}^\ell\frac{\mathrm{d}\mu_{B_\alpha}\,N_{k-\text{gon}}(i)}{\langle AB_\alpha i{-}1i\rangle\langle AB_\alpha ii{+}1\rangle\cdots\langle AB_\alpha i{+}k{-}2,i{+}k{-}1\rangle},
\label{npt1}
\end{equation}
\vspace{-0.15cm}
\end{widetext}
where $[a,b,c,d]$ is the canonical form in $A$ space for the tetrahedron with vertices $Z_a,Z_b,Z_c,Z_d$
\begin{equation}
[a\,b\,c\,d] = \frac{\mathrm{d}\mu_A\, \langle abcd\rangle^3}{\langle Aabc\rangle\langle Aabd\rangle\langle Aacd\rangle\langle Abcd\rangle}.
\end{equation}
The $B_\alpha$ part is a form on the polygon bounded by the lines $(i{-}1\,i)$, $(i\,i{+}1)$,$\dots$, $(i{+}k{-}2,\,i{+}k{-}1)$. The numerator $N_{k-{\rm gon}}$ was given in \cite{Arkani-Hamed:2014dca}. Alternatively, one can triangulate the polygon as a sum of triangles and collect the corresponding differential forms. Note that in the triangulation the $A$-space is given just by simple tetrahedron while the $B$-space is a more complicated polygon.

While for four point case the forms $\Omega_4^{(\ell)}$ and $\widetilde{\Omega}_4^{(\ell)}$ were related for higher points they are different. Following a similar geometric procedure we have to first triangulate the $P$-space; here the general term in the triangulation has six boundaries, and the corresponding space of lines on the $P$-plane has three boundaries for each line. The expression for $\widetilde{\Omega}_n^{(\ell)}$ is then given by the product form
\begin{widetext}
\begin{equation}
\widetilde{\Omega}_n^{(\ell)} = \sum_{i=1}^{n-2}\sum_{j=i{+}1}^{n-1}\sum_{k=j{+}1}^{n} \{i,j,k\} \wedge \prod_{\alpha=1}^\ell\frac{\mathrm{d}\mu_{{\cal L}_\alpha}\,\langle\langle P,i,j,k\rangle\rangle}{\langle (AB)_\alpha i\,i{+}1\rangle\langle (AB)_\alpha j\,j{+}1\rangle\langle (AB)_\alpha k \,k{+}1\rangle},
\label{npt2}
\end{equation}
\end{widetext}
where the bracket $\{i,j,k\}$ is defined by
\begin{equation}
\{i,j,k\}=\frac{\mathrm{d}\mu_P\,\langle\langle P,i,j,k\rangle\rangle}{\langle P\,i\rangle\langle P\,i{+}1\rangle\langle P\,j\rangle \langle P\,j{+}1\rangle\langle P\,k\rangle \langle P\,k{+}1\rangle},
\end{equation}
with 
\begin{align}
\langle \langle P,i,j,k\rangle\rangle &= P^{I_1J_1K_1}  P^{I_2J_2K_2} Z_i^{I_1}Z_{i\psl 1}^{I_2}Z_j^{J_1}Z_{j\psl 1}^{J_2}Z_k^{K_1}Z_{k\psl 1}^{K_2}\nonumber\\
&\hspace{-0.5cm}= \langle (i\,i\psl 1)\,(P)\cap (j\,j\psl1)\,(P)\cap (k\,k\psl1)\rangle. 
\end{align}
Note that $\langle \langle P,i,j,k \rangle \rangle$ is completely symmetric in $i,j,k$, though the representation in terms of four-brackets does not manifest this symmetry. Thinking for convenience dually of $P$ as a point and the $Z_a$ as planes,  $\{i,j,k\}$ is the canonical form of a cube with opposing facets associated with $(Z_i,Z_{i\psl1})$, $(Z_j,Z_{j\psl1})$, $(Z_k,Z_{k\psl1})$.  
 
%%%%%%%%%%%%%%%%%%%
\section{Exceptional efficiency}
%%%%%%%%%%%%%%%%%%%

Given the integrand for the $n$-point, $\ell$-loop MHV integrand, the differential form for the deepest cuts  $\Omega$ and $\widetilde{\Omega}$ can be straightforwardly computed by taking residues. Explicit expressions for the MHV integrand are available in the literature up to ten loops for $n=4$ \cite{Bourjaily:2016evz}, and up to three loops for any $n$ \cite{ArkaniHamed:2010gh}. We have verified our cut up to five loops for $n=4$ and for general $n$ up to three loops. 

As we have stressed, the deepest cut is sensitive to the most complicated topologies for Feynman diagrams and on-shell processes that can contribute to the amplitude. It is interesting to see this more quantitatively at four-points. For the four-point $\ell$-loop integrand the number of dual conformal invariant integrals contributing on the cut can be counted from the Mathematica code provided in \cite{Bourjaily:2016evz} up to $\ell=10$. In the table below we provide the number of topologies; the complete set of integrals also involves permutation over all loop momenta and cycling in external labels. The number of contributing diagrams is the same for both solutions of the cut.

\vspace{0.2cm}

\begin{center}
\begin{tabular}{|c|c|c|c|}
\hline $\ell$ & total $\#$ of topologies & contributing on cut & $\%$ \\ \hline
4 & 8 & 4 & 50 \\ \hline
5 & 34 & 20 & 58.8 \\ \hline
6 & 229 & 146 & 63.8 \\ \hline
7 & 1873 & 1248 & 66.6 \\ \hline
8 & 19 949 & 13 664 & 68.5 \\ \hline
9 & 247 856 & 172 471 & 69.6 \\ \hline
10 & 3 586 145 & 2 530 903 & 70.6 \\ \hline
\end{tabular}
\end{center}

We see the monotonic increase in the percentage of diagrams contributing as a function of $\ell$. We expect this percentage should approach 100$\%$ for $\ell\rightarrow\infty$. 

Our formulas for $\Omega$ and $\widetilde{\Omega}$ are remarkably simple and compact, while the complete loop integrand gets more and more complicated for higher $\ell$. A notable feature of our expressions is their representation as a sum over pieces each having a trivial product structure over loops. However, this simplicity comes at the cost of introducing spurious poles. These are all the poles in $\Omega$ involving only the intersection point $``A"$ rather than the lines $AB_\alpha$. This phenomenon is by now a familiar one in the on-shell approach to scattering amplitudes, but occurs here in a novel setting. Of course the spurious poles cancel non-trivially in the sum. The existence of such a strikingly simple and unusual representation for this cut is completely mysterious from any conventional point of view (Feynman diagrams, all-loop BCFW recursion or Wilson-loops). However, as we stressed even before embarking on any detailed calculation, the existence of such a picture is made almost trivially obvious from the topological picture of the amplituhedron geometry. 

Any analytic comparison with standard local expressions for the cut would have to proceed by algebraically canceling the spurious poles. This immediately leads to an explosion of complexity: while the formula for $\Omega$ has the same form for any $\ell$, when canceling spurious poles the result gets more complicated for higher $\ell$. Even at four points, when all spurious poles are canceled we get
\begin{align*}
\Omega_4^{(\ell)} = \mathrm{d}\mu_A\,{\cal N}_\ell\prod_{k=1}^\ell\frac{\mathrm{d}\mu_{B_\alpha}}{\langle AB_\alpha12\rangle\langle AB_\alpha23\rangle\langle AB_\alpha34\rangle\langle AB_\alpha14\rangle}
\end{align*}
where for $\ell=2,3$ we get up to symmetrization in $B_\alpha$
\begin{align*}
{\cal N}_2 &= (\langle AB_113\rangle\langle AB_224\rangle+\langle AB_213\rangle\langle AB_124\rangle)\\
{\cal N}_3& = \left[\begin{array}{c} \langle A124\rangle^2 \langle AB_113\rangle\langle AB_223\rangle\langle AB_334\rangle \\ + \langle A234\rangle^2\langle AB_112\rangle\langle AB_213\rangle\langle AB_314\rangle\\ +\langle A234\rangle\langle A124\rangle\langle AB_113\rangle\langle AB_213\rangle\langle AB_324\rangle\end{array}\right].
\end{align*}

The expressions get even more complicated if we rewrite the numerators using $\langle AB_\alpha\,i\,i\psl1\rangle$ to match the Feynman integral expansion. In that case ${\cal N}_2$ would be a sum of four terms corresponding to double box integrals while ${\cal N}_3$ would be given by 24 terms corresponding to tennis court diagrams at three loops. The expressions obviously get more complicated at higher loops and the number of terms matches the number of contributing Feynman integrals. In the numerator ${\cal N}_\ell$ all lines ${\cal L}_\alpha$ are completely entangled and there is no product structure. Again, from this point of view the the amazingly simple product form (\ref{Omeg1}) is a total surprise, and without the geometric picture one would never discover it.

\section{Conclusion}

In this letter we studied the deepest cut in the planar ${\cal N}=4$ SYM theory using the new topological definition of the amplituhedron geometry using sign flips, which allowed us to easily find explicit triangulations and concrete expressions for this cut of the $n$-point MHV amplitudes (\ref{npt1}) and (\ref{npt2}). When compared to the Feynman diagram expansion, the deepest cut probes the most complicated diagrams. We expect that for $\ell\rightarrow\infty$ this cut captures some of the essential properties of the full integrand. Indeed, the deepest cut can be used as a new jumping off point for approaching the determination of the full geometry by gradually relaxing the mutual intersection properties in steps \cite{prep}. It should also be possible to compute the deepest cut for all $k$; as with the MHV case, in the topological picture this should again reduce to merely finding a precise characterization of the one-loop amplituhedron geometry, but the associated analogs of our ``tetrahedral" and ``polygonal" regions are expected to be more non-trivial and interesting. 
\smallskip

{\it Acknowledgment:} We thank  Lance Dixon, Enrico Herrmann, Thomas Lam, and Hugh Thomas for useful discussions and comments. N.A-H. is supported by a DOE grant No. SC0009988. C.L. and J.T. are supported by the DOE grant No. DE-SC0009999 and by the funds of University of California.

\end{document}